\documentclass[sigconf]{acmart}
\AtBeginDocument{%
  \providecommand\BibTeX{{%
    \normalfont B\kern-0.5em{\scshape i\kern-0.25em b}\kern-0.8em\TeX}}}

\setcopyright{acmcopyright}
\copyrightyear{2023}
\acmYear{2023}
\acmDOI{XXXXXXX.XXXXXXX}

\acmConference[Conference acronym 'XX]{Make sure to enter the correct
  conference title from your rights confirmation emai}{June 03--05,
  2018}{Woodstock, NY}
\acmPrice{15.00}
\acmISBN{978-1-4503-XXXX-X/18/06}

\begin{document}

\title{Personalized Transformer-based Ranking for e-Commerce at Yandex}

\author{Kirill Khrylchenko}
\affiliation{%
  \institution{Yandex}
  \streetaddress{P.O. Box 1212}
  \city{Moscow}
  \country{Russia}
  \postcode{43017-6221}
}

\author{Alexander Fritzler}
\affiliation{%
  \institution{Yandex}
  \streetaddress{P.O. Box 1212}
  \city{Almaty}
  \country{Kazakhstan}
  \postcode{43017-6221}
}

\renewcommand{\shortauthors}{Khrylchenko, et al.}

\begin{abstract}
Personalizing user experience with high-quality recommendations based on user activity is vital for e-commerce platforms. This is particularly important in scenarios where the user's intent is not explicit, such as on the homepage. Recently, personalized embedding-based systems have significantly improved the quality of recommendations and search in the e-commerce domain. However, most of these works focus on enhancing the retrieval stage. 

In this paper, we demonstrate that features produced by retrieval-focused deep learning models are sub-optimal for ranking stage in e-commerce recommendations. To address this issue, we propose a two-stage training process that fine-tunes two-tower models to achieve optimal ranking performance. We provide a detailed description of our transformer-based two-tower model architecture, which is specifically designed for personalization in e-commerce.

Additionally, we introduce a novel technique for debiasing context in offline models and report significant improvements in ranking performance when using web-search queries for e-commerce recommendations. Our model has been successfully deployed at Yandex, serves millions of users daily, and has delivered strong performance in online A/B testing.
\end{abstract}

\begin{CCSXML}
<ccs2012>
<concept>
<concept_id>10002951.10003317</concept_id>
<concept_desc>Information systems~Information retrieval</concept_desc>
<concept_significance>500</concept_significance>
</concept>
<concept>
<concept_id>10002951.10003317.10003338.10003343</concept_id>
<concept_desc>Information systems~Learning to rank</concept_desc>
<concept_significance>500</concept_significance>
</concept>
<concept>
<concept_id>10002951.10003260.10003261.10003271</concept_id>
<concept_desc>Information systems~Personalization</concept_desc>
<concept_significance>500</concept_significance>
</concept>
<concept>
<concept_id>10002951.10003317.10003347.10003350</concept_id>
<concept_desc>Information systems~Recommender systems</concept_desc>
<concept_significance>500</concept_significance>
</concept>
<concept>
<concept_id>10010147.10010257.10010282.10010292</concept_id>
<concept_desc>Computing methodologies~Learning from implicit feedback</concept_desc>
<concept_significance>500</concept_significance>
</concept>
<concept>
<concept_id>10010147.10010257.10010293.10010294</concept_id>
<concept_desc>Computing methodologies~Neural networks</concept_desc>
<concept_significance>500</concept_significance>
</concept>
</ccs2012>
\end{CCSXML}

\ccsdesc[500]{Information systems~Information retrieval}
\ccsdesc[500]{Information systems~Learning to rank}
\ccsdesc[500]{Information systems~Personalization}
\ccsdesc[500]{Information systems~Recommender systems}
\ccsdesc[500]{Computing methodologies~Learning from implicit feedback}
\ccsdesc[500]{Computing methodologies~Neural networks}

\keywords{deep learning, personalization, recommender systems, e-commerce, learning-to-rank, debiasing}


\maketitle

\section{Introduction}
E-commerce platforms have become an essential part of our society, allowing us to exploit and explore our interests through access to a wide variety of items. Recommender systems aim to alleviate information overload and present relevant and interesting results to users from billions of potential choices. 

Recently, many large e-commerce platforms (e.g. Amazon~\cite{amazon1, amazon2}, Taobao \cite{alibaba1, alibaba2, alibaba3}, eBay \cite{ebay1, ebay2}, Walmart \cite{walmart1, walmart2, walmart3}, JD.com \cite{jd1, jd2}, Facebook Marketplace \cite{meta1, meta2}) have integrated deep learning into their recommender systems and search engines. However, most of these works focus on the retrieval stage, which produces a small set of candidates for further filtering and ranking by subsequent stages. Similarity features produced by embedding-based retrieval (EBR) models can also improve ranker models. For example, \citet{meta1} report +1.24\% online in-conversation uplift in Que2Search from using two-tower cosine similarity feature trained with retrieval loss. 

Although some models attempt to bridge the gap between retrieval and ranking \cite{alibaba3, meta2, alibaba3}, we argue that retrieval and ranking still vary significantly. Retrieval is typically concerned with recall optimization and heavily utilizes weakly supervised signals like clicks. Ranking is based on sparser signals, such as purchases. 

In this paper, we introduce a ranking-oriented transformer-based two-tower model for personalization. The user-item similarity features generated with our approach greatly improve the quality of subsequent ranking models. Our key contributions are as follows:
\begin{itemize}
    \item We present an inductive transformer-based two-tower model architecture that is specifically optimized for the personalized ranking of e-commerce recommendations. To the best of our knowledge, this is the first work to focus on ranking-oriented personalization transformers for recommendations in the e-commerce domain.
    \item We propose a novel two-stage training technique that combines recall-oriented pre-training with ranking-oriented fine-tuning. This approach results in significant improvements in ranking quality and addresses the challenge of sparse positive ranking signals.
    \item We show that integrating web-search history into e-commerce recommendations significantly improves ranking quality.
    \item We define the concept of context bias and propose a context debiasing technique to remove context-related signal from offline, non real-time models. 
    \item Our model has been successfully deployed in an industrial setting and has demonstrated strong performance improvements on the homepage and cart landing page of the large e-commerce platform Yandex Market through online A/B testing.
\end{itemize}

\section{Related Works}
To our knowledge, the first two-tower model was proposed by Microsoft: DSSM \cite{dssm} generated query and document embeddings for the web-search domain. Many companies have since adopted two-tower models for various information retrieval tasks such as recommendations, search, and ads. Information retrieval involves a multi-stage process of information filtering: billion-sized catalogs are filtered down to a few candidates to be presented to the user. The first stage, called retrieval, is where two-tower models are most commonly used. EBR systems usually constitute one of multiple retrieval channels~\cite{meta1, amazon1, alibaba3, walmart1}. Another popular choice is the inverted index with boolean matching based on text or visual signals~\cite{msuru, groknet, make1, make2}. While inverted indices have good relevance due to exact term matching, EBR is tolerant to misspellings and allows for personalization and context awareness. EBR models are typically trained to optimize softmax loss with in-batch or uniformly sampled negatives. \citet{google1} used importance sampling to correct sampling bias of in-batch negatives for Youtube. \citet{google2} introduced further improvements for retrieval using mixed negative sampling for Google App Store. Recently, there has been a trend to increase consistency between the ranking and retrieval stages with auxiliary losses~\cite{meta2, alibaba3, jd2}.

The ranking stage is usually one of the last stages in information filtering, where compute-intensive models are used on hundreds or thousands of candidates. Deep learning models for personalized ranking mostly follow the Embedding\&MLP paradigm proposed by \citet{wide&deep}. Such models tend to abandon two-tower architecture in favor of fusing user, context, and target item information. YoutubeDNN~\cite{youtubednn} formed user video-watching and searching activity vectors by averaging historical user event embeddings. DIN~\cite{din} used a convex combination of historical user event embeddings, essentially an attention layer with the target item as query and historical events as keys and values. DIEN~\cite{dien} and BST~\cite{alibaba1} improved on DIN with GRU and transformers. However, these models use an early fusion of the target item and user history items, which is very costly in production even for the re-ranking stage and limits our ability to use batch serving. This motivates us to explore two-tower models. There are works that employ transformer-based two-tower models to produce similarity-based features for ranking in search \cite{amazongraph, baidu} and ads \cite{uniretriever}. However, these models are not personalized and only use query and document information.

Personalization with sequential recommenders based on user histories is largely associated with the next-item-prediction (NIP) task, which essentially turns recommendation into a language modeling problem. First works in this field were motivated by the progress in natural language processing, e.g. ELMo~\cite{elmo}, Transformer~\cite{transformer}, GPT~\cite{gpt}, BERT~\cite{bert}. GRU4Rec \cite{gru4rec}, Caser~\cite{caser} and SASRec~\cite{sasrec} were the pioneer works to model the NIP task with GRU, CNNs, and transformers respectively. Bert4Rec~\cite{bert4rec} trained a recommender in a BERT-like bidirectional fashion with MLM. CL4Rec \cite{cl4rec} used contrastive learning for pre-training, and CARCA \cite{carca} promoted inductive context-aware models with content-based item embeddings. However, most of these works conduct evaluation on academic datasets with questionable practices such as sampled negatives~\cite{nipsampled} and absence of a time-based train-test split~\cite{niptime1, niptime2}. Furthermore, the NIP task is subject to selection bias which encourages models to imitate the logging policy~\cite{niplogging}. Additionally, due to the scale of academic datasets, these works are prone to unnecessary or even harmful inductive bias: customized model architectures, auxiliary losses, and special pre-training regimes that do not benefit real-world web-scale recommendations. Large companies also conduct research and develop sequential recommenders: Pinterest~\cite{pinterest2}, Alibaba~\cite{alibaba1, alibaba2}, eBay~\cite{watchlist, ebay1, ebay2}, NAVER~\cite{clue, reclm}, Etsy~\cite{etsyads}, Spotify~\cite{spotify}, etc. They usually provide correct evaluation schemes, use web-scale datasets, and sometimes provide results of online A/B testing.

To combat the feedback loop, it's important to mitigate the effects of various recommendation biases like position, selection, examination, trust, and popularity bias. There are several debiasing approaches: (1)~reweighting samples based on inversed propensity scores~\cite{ipw1, ipw2}; and (2)~modeling bias explicitly with a separate model~\cite{pal, watchnext, airbnb, disentanglement}, detaching the bias model during deployment. Motivated by the second approach, we introduce the concept of context bias and propose a context debiasing technique to improve the ranking quality of offline models.

Although enriching user e-commerce history with web-search queries is a case of cross-domain recommendations (CDR), we do not delve into CDR research because it is not a goal of our work to explore and improve upon various CDR methods. To the best of our knowledge, the most notable works that improve the quality of e-commerce recommendations with web-search data are from NAVER~\cite{clue, reclm}.

\section{Modeling}
In this section, we describe our two-tower model architecture, shown in Figure \ref{fig:model}. We discuss the item tower, user tower, and similarity function separately, before introducing an additional context tower for context debiasing. Finally, we present our two-stage training regime, which involves pre-training the model for retrieval-oriented tasks and then fine-tuning it for ranking-focused tasks.
\subsection{Item Tower} \label{sec:item}

    \begin{figure*}[h]
  \centering
  \includegraphics[width=0.9\textwidth]{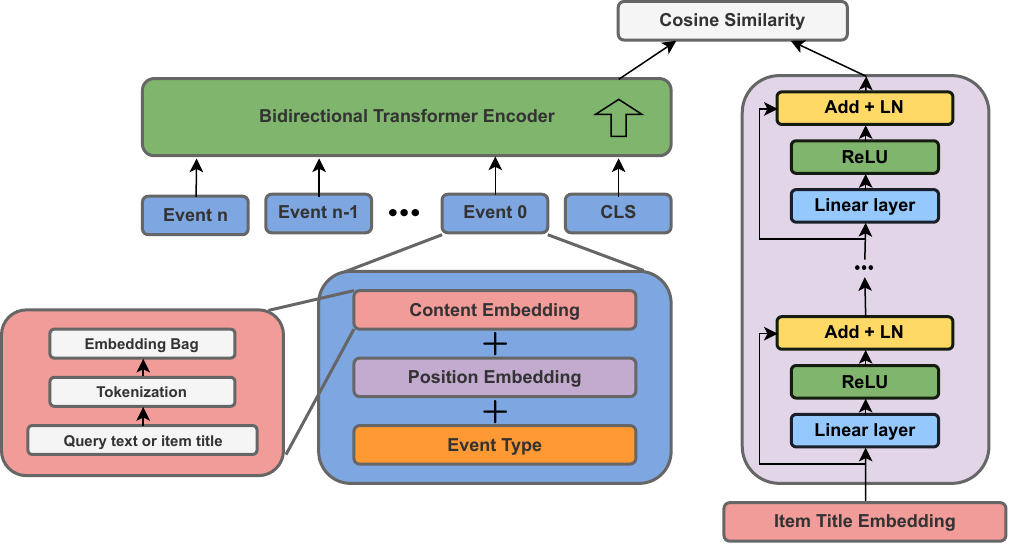}
  \caption{Model architecture. Add+ln denotes $\text{layernorm}(x + y)$.}
    \label{fig:model}
\end{figure*}
    In the e-commerce domain we are tasked with recommending products to the users. Unlike some domains such as web-search, products provide a rich source of information including category, title, description, structured attributes (e.g. brand, color, flavor, material), images, and reviews. Product catalogs are very dynamic with some items available in a single quantity and new items arriving daily, making it important to consider the cold-start and distribution shift problems. To increase generalization, avoid memorization and alleviate cold-start we do not use identifier-based learnable embeddings. Instead, we employ content-based item representations using only titles.

    Product titles are processed as CBOW~\cite{cbow} using a wordpiece~\cite{wordpiece} tokenizer built on web-search data with a vocabulary of 103295 tokens. Token embeddings are summed which does not degrade performance compared to averaging.

    The item tower architecture was motivated by our experience with personalization transformer-based models for web search and display advertising. Initially, in web-search we used a cross-encoder model with user and item features fused in the same transformer encoder. Later on we switched to two-tower models to handle millions of RPS in display advertising. The quality of the model decreased significantly when we used a Multi-Layer Perceptron (MLP) as an item tower. However, using transformer encoder as an item tower on top of products represented as a single CBOW title embedding produced good results.
    
   Simplifying the transformer encoder with a single input embedding lead to the architecture presented on the right in figure \ref{fig:model}. This architecture differs from the MLP in that it includes residual connections and layer normalizations, similar to those found in a transformer:
    \begin{displaymath}
        x_{k+1} = \text{LayerNorm}_k(\text{ReLU}(\text{Linear}_k (x_k)) + x_k)
    \end{displaymath}
    where $k$ is a layer number and $x_1~\in~\mathbb{R}^d$ is an input title embedding.
    
    With such architecture, we were able to preserve most of the initial cross-encoder quality gains. Incorporating contrastive learning for pre-training further improved the model quality compared to the cross-encoder.

    Following the item tower, the item embedding is $l_2$ normalized to unit length. 

\subsection{User Tower}
    To our knowledge, when choosing which types of user events to use in a sequence encoded by a transformer, the first best guess is to use the same types of events you are trying to predict. We form a chronologically ordered sequence from the user's clicks, add-to-carts, add-to-favorites, and purchases. 
    
    We also observed quality gains from enriching user histories with web-search queries. Interestingly, using web-search document clicks did not lead to any further improvements. We explore two ways of incorporating web-search history: using a separate transformer encoder on a fixed number of the most recent user's web-search queries and fusing web-search information with e-commerce activity into a single event sequence. Moreover, we believe that combining diverse data types, such as queries and products, yields even better results in an inductive setting that utilizes content-based event representations.

    Three types of embeddings are summed up to form an event embedding:
    \begin{itemize}
        \item \textbf{Content embedding.} Queries and product titles are tokenized using the wordpiece tokenizer from section \ref{sec:item}. The embedding matrix is shared across queries and products.
        \item \textbf{Positional embedding.} Each event is assigned a chronologically reversed absolute position: the latest event gets position $0$, the second latest gets position $1$, and so on. Position embeddings are learnable.
        \item \textbf{Event type embedding.} Every event type, including click, add-to-cart, add-to-favourites, purchase, web-search query, has a learnable embedding.
    \end{itemize}
    Together with learnable [CLS] embedding, event embeddings are fed to a bi-directional transformer encoder. A post-transformer contextualized [CLS] embedding is $l_2$ normalized and used as a final user representation. We use a standard transformer encoder architecture \cite{transformer, bert} with post-normalization. All embeddings, except for [CLS], are layer-normalized prior to the transformer.
    
    Similar to Pinnerformer \cite{pinterest2}, we opt to use a single user embedding to simplify maintenance and deployment. Our model is used in a batch serving scenario with user and item embeddings being recalculated daily. Multiple user embeddings would result in a linearly scaled memory cost for KV-storage and complicate integration into downstream applications. However, using multiple user embeddings usually produces better quality and helps to enforce diversity, which is why we usually employ multiple embeddings for real-time models.

    We use a bi-directional transformer encoder without causal masking and teacher forcing, only the [CLS] representation is used to produces scores.

\subsection{Similarity function}
We use inner product as a user-item similarity function. Since we $l_2$ normalize embeddings, it is equivalent to cosine similarity:
\begin{displaymath}
    r_{ui} = \frac{\left\langle v_u, v_i \right\rangle}{\parallel v_u \parallel \parallel v_i \parallel}
\end{displaymath}
where $v_u, v_i \in \mathbb{R}^d$ are user $u$ and item $i$ embeddings. In our experiments, we found that normalizing embeddings usually results in a more stable training process.

\subsection{Context Debiasing}
    When a recommendation request comes, typically a variety of contextual data, such as device, operational system, time of day, seed item, search query, is available along with user and item information. Using context information in real-time models is paramount and usually brings significant improvements.

    We will discuss deployment in more detail in Section \ref{sec:deploy}, but it's worth noting that we deploy our model in batch-serving mode, i.e. user and item embeddings are recalculated on a daily basis. As a result, we do not have access to context information during model inference. Since the context has very useful signal, our models tend to implicitly infer current user context from history. However, it is unnecessary to spend model and user embedding capacity for implicit context information. We expect that our gradient boosting trees ranker model (e.g. Catboost~\cite{catboost}), for which we are constructing a similarity-based feature, already utilizes context information efficiently. Additionally, our data is skewed towards certain user contexts that are more prevalent in real-world scenarios, which can impact our performance for less popular contexts.
    
    Inspired by debiasing techniques aimed at mitigating effects from various biases, we introduce context debiasing technique for alleviating redundancy of context information in our ranker. Without context debiasing, our model estimates the following probability:
    \begin{displaymath}
        \text{P} (i\vert u) = \sum_{ctx \in C} \text{P} (i \vert u, ctx) \text{P} (ctx \vert u)
    \end{displaymath}
    where $\text{P} (i\vert u)$ is a probability that the user $u$ clicks on item $i$, and $C$ is a set of all possible contexts.

    Instead, during training we learn a separate context tower using actual user context:
    \begin{displaymath}
        \text{P} (i\vert u, ctx) = \sigma (r_{ui} + r_{ctx})
    \end{displaymath}
    where $r_{ui}$ is learned similarity between user $u$ and item $i$, and $r_{ctx}$ is a learnable scalar based on a given context $ctx$. 

    During deployment, we detach the context tower and use only user-item similarity. A possible modification to this technique would be to fuse context with item and user, for example, by forming a context embedding and calculating the inner product between context and item. However, in our experiments, we didn't see any improvements with this approach.

    Similar to \citet{disentanglement}, we attempted to apply dropout to the output of the context tower but it did not lead to any quality improvements. This could be related to the fact that our context tower has the simplest possible form. We do not exclude the possibility that dropout will be useful with a more complex context tower.

\subsection{Two-stage training}
    The most common loss function for deep learning two-tower personalization models is a sampled softmax loss with random and in-batch negatives, which essentially makes the model retrieval-focused. Negative implicit feedback based on impressions is rarely used, because: (1) using logs with non-positive impressions dramatically increases pipeline complexity; (2) such models perform poorly on retrieval tasks due to sample selection bias; (3) all impressed items are very similar \cite{ebay1} and examination bias leading to very noisy data with large amounts of false negatives. 
    
    However, the sample selection bias problem goes both ways. When trained with sampled negatives, models underperform on ranking scenarios which present much harder impressed negatives. We demonstrate that tuning the model directly on a ranking scenario with pairwise ranking loss and impressed negatives results in significant gains.

    Still, due to the data sparsity of positive feedback (such as purchases), training large two-tower models on ranking from scratch yields almost zero results. This issue can be mitigated with transfer learning. We propose a two-stage training scheme for training ranking-oriented deep learning two-tower models.

    \subsubsection{Pre-training stage} In the pre-training stage, the model is trained in a standard retrieval-oriented regime with sampled softmax loss function:
    \begin{displaymath}
    \mathcal{L}_{pretrain}(u, p, N) = - \log \frac{\exp(\tau \cdot r_{up})}{\exp(\tau \cdot r_{up}) + \sum_{n \in N} \exp(\tau \cdot r_{un})}
    \end{displaymath}
    where $\tau$ represents the temperature parameter, $r_{ui}$ denotes the similarity between user $u$ and item $i$, $p$ corresponds to a positive item, and $N$ is a collection of in-batch negatives. Along with on-device in-batch negatives, we utilize item embeddings from all other GPU workers as negative samples. User clicks, add-to-carts, add-to-favourites and purchases are used as positive item interactions.

    \subsubsection{Fine-tuning stage} During the fine-tuning stage, we use a separate loss for each type of the positive impressed signal, including clicks, add-to-cart, add-to-favorites, and purchases. To ensure that our similarity scores remain well-calibrated during continuous training, we employ a combination of pairwise and pointwise losses.

    Initially, we used a separate pointwise loss for each type of the positive signal, but later we switched to a single pointwise loss based on clicks:
    \begin{displaymath}
        \text{BCE}_{\text{click}}(u, i) = -y_i \log f_{ui} - (1 - y_i) \log(1 - f_{ui})
    \end{displaymath}
    where $y_i$ indicates whether item $i$ is clicked, $f_{ui}:=\sigma(\alpha_{cl} r_{ui}+\alpha_{ctx} r_{ctx}+\beta_{cl})$ is the predicted probability of item $i$ being clicked by user $u$, and $\alpha_{cl}, \alpha_{ctx}, \beta_{cl}$~are learned scalar parameters. $r_{ctx}$ is the context tower prediction.

    The initial pairwise loss for each type of positive signal had a form:
    \begin{equation}
    \label{eq:pairwise}
    \begin{split}
    \text{BPR}_{k}(u, p, n) &= -\log \sigma (\gamma_{k} (r_{up} - r_{un})) = \\
    &= -\log \frac{e^{\gamma_k r_{up}}}{e^{\gamma_k r_{up}}+ e^{\gamma_k r_{un}}}
    \end{split}
    \end{equation}
    where  $k$ denotes a type of positive feedback, $u,p,n$ represent user, positive and negative items respectively, and $\gamma_k$ is a learned scalar.
    
    We adopted a solution from \citet{calibrated} to pairwise losses, replacing the exponents with sigmoids in equation \ref{eq:pairwise}. With this approach, we achieve well-aligned ranking and regression objectives. Furthermore, sigmoids allow us to integrate context debiasing into pairwise losses. Our final pairwise ranking loss is:
    \begin{displaymath}
     \text{BPR}_{k}(u, p, n) = -\log \frac{f_{up}}{f_{up} + f_{un}}
    \end{displaymath}
    where $f_{ui} := \sigma \left(\gamma_k r_{ui} + \gamma_{k,ctx} r_{ctx} + \beta_k\right)$, with additional learned scalars $\gamma_{k, ctx}, \beta_k$.
    
    Our final objective for fine-tuning the model is shown below. For brevity, we omit specific arguments of each loss component:
    \begin{equation}\label{eq:finetune}
        \mathcal{L}_{finetune} = \text{BPR}_{click} + \text{BPR}_{cart} + \text{BPR}_{fvrt} + \text{BPR}_{prch} + \text{BCE}_{\text{click}}
    \end{equation}
    where the pairwise loss is calculated for each targetwise ordered item pair within a recommendation request and pintwise loss is calculated for each impressed item.
\section{Experiments}
\subsection{Dataset}
We gather one year of logs from Yandex e-commerce platform Yandex Market for training, including clicks, add-to-cart, add-to-favorites, purchases, and web queries. For fine-tuning, we collect impressions from all recommendation surfaces at Yandex Market. Due to the proprietary nature of the data, we don't disclose dataset sizes. 

 For each positive user interaction, a pre-training sample is formed from the user history up to that interaction with a one-day delay. Such delay for history events is important because (1) it is consistent with the production daily batch serving job scenario, and (2) setting a low history delay may lead to an overly easy pre-training task. A single item usually generates a sequence of positive interactions within a short time period, e.g. click~$\to$~add-to-cart~$\to$~purchase, which is why it becomes easy to predict add-to-carts based on clicks and predict purchases based on add-to-carts with in-batch negatives and short history delay. Unlike the fine-tuning stage, pre-training also utilizes organic positive interactions produced by the user outside of recommendations.

For fine-tuning, we utilize impressions from all recommendation surfaces on Yandex Market. We group items impressed together and filter out groups without any positive interactions. An impression is considered positive if the impressed item had a positive interaction with the user within a short period following the recommendation. To speed up training and simplify the implementation of pairwise losses, we pack together all impressed items from the same recommendation surface into a single dataset record.

Although almost every recommendation surface has its own GBT ranker, for brevity we provide results of offline evaluation only for the most user-centric surfaces: retargeting and discovery. Retargeting is an unconstrained form of personalization akin to eBay's Recently Viewed Items \cite{ebay1} module. Discovery limits recommended items to previously undiscovered ones. We use the next ten days after the training period to evaluate our model as a feature in the GBT ranker. As a metric, we report a test relative nDCG gain from including our similarity feature in the GBT feature set.

Context debiasing tower uses recommendation surface identifier and user device as input features. The context tower produces a sum of two learned scalar values, for surface and device respectively.

\subsection{Implementation Details}
We use quite a large amount of user events --- 1024 latest user events. The transformer encoder has four layers with hidden size 256 and four attention heads. The candidate tower has four layers with hidden size 1024. The temperature parameter for softmax loss is learnable. We split embedding matrices, transformer, candidate tower, and loss parameters into separate parameter groups with groupwise gradient norm clipping and differently tuned learning rates. Learning rate is warmuped for 2500 steps and then linearly decays till the end of the training. Pre-training is done on 16 A100 40g hosts with an effective batch-size of 2048 while fine-tuning is done on 8 A100 40g hosts with an effective batch size of 4096. Pre-training is done for three epochs and fine-tuning takes a single epoch due to overfitting. Training is done in a multi-host distributed setting with PyTorch~\cite{pytorch} and Deepspeed~\cite{deepspeed}.

We continuously fine-tune the model on new chunks of data, weekly. During continuous training, we freeze sigmoid inner parameters ($\alpha_k, \beta_k$). Also, we use a constant learning rate for all parameter groups. Only second training stage, presented in equation \ref{eq:finetune}, is employed. Optimizer state is not reused across iterations.

\subsection{Offline Experimental Results}
\begin{table}
  \caption{Incremental Model Improvements (Relative nDCG Improvement).}
  \label{tab:offline}
  \begin{tabular}{llcc}
    \toprule
    & Model & Retargeting & Discovery \\
    \midrule
    1& Sampled Softmax Loss & +0.233\% & +0.882\%  \\
    2& Two-stage Training & +0.476\% & +1.377\% \\
    3& Calibrated Pairwise Ranking & +0.501\% & +1.367\% \\
    4& Retaining Single Pointwise Loss & +0.521\% & +1.405\% \\
    5& History Length (256 to 512) & +0.541\% & +1.588\% \\
    6& Pointwise Loss Weight (1.0 to 0.1) & +0.566\% & +1.631\% \\
    7& Context Debiasing & +0.586\% & +1.645\% \\
    8& Web-Search Queries & +0.677\% & \textbf{+2.061\%} \\
    9& History Length (512 to 1024) & \textbf{+0.755\%} & +1.981\% \\
  \bottomrule
\end{tabular}
\end{table}
The main experimental results are presented in Table \ref{tab:offline} as a sequence of incremental improvements. The baseline (L1) is only trained on the retrieval task. Fine-tuning on the pairwise ranking task (L2) brings significant improvements. Calibrated ranking (L3) bridges the gap between pointwise and pairwise losses. Removing all pointwise losses, with the exception of clickwise loss (L4), demonstrates an uplift in quality in both retargeting and discovery scenarios. Scaling history length to 512 (L5) greatly improves discovery. We then reduce the importance of pointwise loss (L6) and apply context debiasing (L7). Finally, we enrich user histories with web-search queries. First, we keep the same maximum history length as in previous measurements (L8) and then we improve on retargeting by expanding the history length to 1024 events (L9).

\begin{table}
  \caption{Comparison of Training Regimes for Two-Stage Training (Relative nDCG Improvement).}
  \label{tab:pretrain}
  \begin{tabular}{cccc}
    \toprule
    Pre-training & Fine-tuning & Retargeting & Discovery \\
    \midrule
    - & \checkmark & +0.031\% & +0.033\% \\
    \checkmark  & - & +0.233\% & +0.882\%  \\
    \checkmark  & \checkmark  & \textbf{+0.476\%} & \textbf{+1.377\%} \\
  \bottomrule
\end{tabular}
\end{table}
The benefits of two-stage training are illustrated in Table~\ref{tab:pretrain}. The sparsity of positive feedback in the pairwise ranking scenario renders the one-stage training approach ineffective, whereas pre-training on the retrieval task mitigates this effect.

\begin{table}
  \caption{Impact of Enriching User History with Web-Search Queries (Relative nDCG Improvement).}
  \label{tab:web}
  \begin{tabular}{lclcc}
    \toprule
    & Fusion & History (ecom + web)& Retargeting & Discovery \\
    \midrule
    1& - & 512 + 512 & +0.704\% & +1.787\%  \\
    2& \checkmark & 512 & +0.677\% & \textbf{+2.061\%} \\
    3& \checkmark & 1024 & \textbf{+0.755\%} & +1.981\% \\
  \bottomrule
\end{tabular}
\end{table}
Enhancing user history with web-search queries works better with an early fusion of e-commerce and web-search data, which is demonstrated in the Table \ref{tab:web}. When mergin history (L2, L3), we take the chronologically latest user events across both web-search and e-commerce activity. Without fusion (L1), we take 512 latest e-commerce events and 512 latest web-search events and train separate transformer encoders to form two embeddings for each user. Single scalar prediction value is calculated by summing up the scaled inner products for each of the user embeddings with an item embedding. During evaluation, we form two distinct features, separately calculating user-item similarities for web-search and e-commerce user representations. Although comparing single feature to multiple features is unfair, fusing information still works better.

\subsection{Online Deployment}\label{sec:deploy}
User and item embeddings are recalculated daily with a batch serving job. Once recomputed, the user embeddings are uploaded to kv-storage. During serving, the item embeddings are stored in RAM as a hash table and embedding lookup is performed when necessary.

We report the number of billed orders that are attributed to recommendations. If an item has been impressed to a user within a short period before the item's order, the item's order is attributed to recommendations. An order as a whole is attributed to recommendations if at least one item in the order is attributed to recommendations. An order is marked as new if it has items that the user has never encountered on our platform before the recommendation.

\begin{table}
  \caption{Online A/B Test Results.}
  \label{tab:online}
  \begin{tabular}{lll}
    \toprule
    Surface &\#RecOrders&\#New RecOrders\\
    \midrule
    Homepage & +6\% & +10\% \\
    Cart page & +5\% & +7\%\\
  \bottomrule
\end{tabular}
\end{table}
Table \ref{tab:online} shows that our model increased the number of orders attributed to recommendations by 6\% on the homepage and by 5\% on the cart page. New orders attributed to recommendations increase by +10\% on the homepage and by +7\% on the cart page. Furthermore, we increased the time spent on the next visited page after the homepage by +1.5\%.

Our online A/B test revealed an effect similar to Pinterest~\cite{decay}: the uplift from our model was substantially higher in the first few days compared to the rest of the A/B test.

\section{Conclusion}
In this paper, we demonstrated the importance of ranking-focused fine-tuning of deep learning embedding-based personalization models. We proposed a two-stage training procedure with retrieval-oriented pre-training and ranking-oriented fine-tuning and introduced context debiasing to improve offline models. Enrichment of user e-commerce histories with web-search data demonstrated further improvements. Lastly, we described in detail our two-tower transformer-based personalization model for e-commerce which was validated through offline and online experiments. The described model has been deployed at the Yandex e-commerce platform Yandex Market to serve main app traffic.

We are currently working on a multi-domain real-time context-aware two-tower transformer-based model for both e-commerce recommendations and search at Yandex Market. We also explore ways to improve item embeddings with (1) item transformers, (2) additional product data like structured attributes, descriptions, and product images, and (3) graph neural networks.

\begin{acks}
We would like to express our gratitude to the team at Yandex Market for their support during the research and deployment of this model. In particular, we would like to thank Mikhail Denisenko for his significant contribution to this project. We are also grateful to Artur Ilichev and Ivan Lapitsky for their assistance. Additionally, we want to acknowledge the rest of our Recsys team, including Ivan Guschenko-Cheverda, Vsevolod Svetlov, and Sergey Ovcharenko. Without their hard work and dedication, this research would not have been possible.
\end{acks}

\bibliographystyle{ACM-Reference-Format}
\bibliography{references}


%% \section{Research Methods}

%% \subsection{Part One}

\end{document}